\documentclass[a4paper]{article}

\usepackage[margin=1in]{geometry} % full-width

\usepackage{amsmath}
\usepackage{amsthm}
\usepackage{amssymb}

\usepackage[utf8]{inputenc}
\usepackage{hyperref}

\usepackage{graphicx, color}
\usepackage{mathrsfs} % for \mathscr command

\usepackage{lipsum}
\usepackage[utf8]{inputenc}
\usepackage[T1]{fontenc}
\usepackage{comment}
\usepackage{multirow}
\usepackage{longtable}
\usepackage{booktabs}
\usepackage{lscape}

\usepackage[sort&compress,numbers,square]{natbib}

\title{Cryptocurrency co-investment network: \\ token returns reflect investment patterns}

\author{Luca Mungo $^{1,2, *}$, Silvia Bartolucci$^3$, Laura Alessandretti $^4$}

\date{ \small $^1$Mathematical Institute, University of Oxford, Andrew Wiles Building, Woodstock Rd, Oxford OX2 6GG. \\ $^2$Institute for New Economic Thinking at the Oxford Martin School, Manor Road Building, Oxford OX1 3UQ. \\ $^3$ Dept. of Computer Science, University College London, 66-72 Gower Street, WC1E 6EA London (UK) \\ $^4$ Technical University of Denmark, DK-2800 Kgs., Lyngby, Denmark \\ $^*$ luca.mungo@maths.ox.ac.uk\\
}

\begin{document}

\flushbottom
\maketitle

\begin{abstract}
    Since the introduction of Bitcoin in 2009, the dramatic and unsteady evolution of the cryptocurrency market has also been driven by large investments by traditional and cryptocurrency-focused hedge funds.  
    Notwithstanding their critical role, our understanding of the relationship between institutional investments and the evolution of the cryptocurrency market has remained limited, also due to the lack of comprehensive data describing investments over time. 
    In this study, we present a quantitative study of cryptocurrency institutional investments based on a dataset collected for $1324$ currencies in the period between $2014$ and $2022$ from Crunchbase, one of the largest platforms gathering business information.
    We show that the evolution of the cryptocurrency market capitalization is highly correlated with the size of institutional investments, thus confirming their important role. 
    Further, we find that the market is dominated by the presence of a group of prominent investors who tend to specialise by focusing on particular technologies. 
    Finally, studying the co-investment network of currencies that share common investors, we show that assets with shared investors tend to be characterized by similar market behavior.  
    Our work sheds light on the role played by institutional investors and provides a basis for further research on their influence in the cryptocurrency ecosystem.
\end{abstract}

% Main Body starts here

\section{Introduction}
\label{sec:introduction}

Since the introduction of Bitcoin in 2009 \cite{Satoshi2008}, the cryptocurrency market has experienced bewildering growth, surpassing an overall market capitalization of 1 trillion dollars in early 2021. Beyond private investors, the development of the market was fostered and nurtured by cryptocurrency hedge funds and Venture Capital (VC) funds, with institutional investments in cryptocurrency-related projects reaching an estimated amount of $17$ billion US dollars in 2021 \cite{Bloomberg, fidelityreport}. 

A growing number of traditional financial firms and hedge funds in Europe and the U.S. are also exploring avenues for investments in cryptocurrency via different channels, including, but not limited to, through including cryptocurrency into their portfolios, investing through tokenisation in equity of blockchain companies, and exploiting more regulated tools such as crypto futures, options, and ETFs \cite{OECD,fidelityreport}. Unfriendly regulations, high volatility, and lack of reliable valuation tools, amongst other issues, have so far hindered widespread adoption and institutionalisation of these assets \cite{rauchs20192nd,gurguc2018cryptocurrencies,fidelityreport}. Most crypto platforms, for instance, lack regulatory and supervisory oversight concerning trading, disclosure, anti-money laundering, and consumer protection measures, forming what has also been described as a ``shadow financial system'' \cite{auer2022banking}. Nonetheless, recent challenging events affecting the economy and markets, i.e., the U.S. elections, Brexit in Europe, and the global pandemic, have gradually accelerated the uptake \cite{fidelityreport}. 

Despite these developments, the effects of institutional investments on the cryptocurrency market are still little understood, also due to the lack of comprehensive quantitative data.

A growing body of literature has so far focused on the properties of the rapidly evolving crypto market ecosystem, shedding light on critical aspects such as market efficiency \cite{sigaki2019,vidal2018semi}, asset pricing bubbles \cite{chen2019sentiment}, the dynamics of competition between currencies \cite{Dowd, Luther}, and the impact of collective attention \cite{elbahrawy2019wikipedia}. Given the digital and decentralised nature of crypto assets, a major focus has been to understand the drivers of price fluctuations and how to properly value these assets. Studies using empirical data have focused on understanding the price dynamics using machine learning techniques \cite{alessandretti2018machine, walther, elbahrawy2019wikipedia, mcnally2018predicting, chen2020bitcoin, akyildirim2020prediction}, also including socio-economic signals (e.g., sentiment gathered from social media platforms) that appears to be intertwined with the price dynamics \cite{garcia2014digital,sentimentaste, ortu2022technical, lucchini_et_al_2020}.Research has also shown that movements in the market can be tied to macroeconomic indicators, media exposure and public interest \cite{lyocsa2020impact, corbet2020impact}, policies and regulations \cite{borri2020regulation}, and indeed the behavior of other financial assets \cite{nguyen2022correlation}.

In the context of institutional investments, the recent growing interest in mixed portfolios of crypto and traditional assets\cite{OECD} has paved the way to research looking at optimal portfolio allocation. Studies have focused on the composition of mixed portfolios,i.e., including traditional (bonds, commodities, etc.) and crypto assets \cite{koutsouri2020balancing, platanakis2020should}, and crypto-only portfolios \cite{hu2019modelling,ahelegbey2021crypto} testing the performances of different allocation and re-balancing strategies. It was suggested that the participation of institutional investors in both crypto and traditional markets might lead to potential spillovers and increased contagion risks between traditional finance and decentralised finance (DeFi) \cite{OECD}. 

Understanding the behavior of institutional investors and its effect on the structure and evolution of the cryptocurrency markets is therefore of paramount importance to quantify the mutual impact between DeFi and traditional entrepreneurial finance \cite{OECD,shakhnov2020r}. So far, most of the research available is based on qualitative surveys by private companies of investors in Europe and U.S., which aim to identify market trends and issues, e.g., barriers to adoption and current channels to exposure in crypto \cite{fidelityreport,OECD}. In Sun, 2021 \cite{sun2021factors}, for instance, the authors surveyed 33 Asian firms to investigate whether price volatility lowers institutional investors’ confidence and to quantify the role played by the familiarity of investors with the technology in the selection of crypto assets. In \cite{ciaian2022environmental} the authors analysed the connection between investors’ ESG preferences and crypto investments exposure using household-level portfolio data gathered from the Austrian Survey of Financial Literacy (ASFL). The analysis suggests that crypto investments are more strongly driven by social and ethical preferences compared to traditional investments (e.g., bonds).

In Liu, 2021, \cite{liu2021} the authors provide a first quantitative exploration of the investor's network focusing on data for investments on $\sim 300$ ERC-20 tokens. Their analysis shows that less central tokens in the investment network have also low market capitalization and trading volume, poor liquidity, and high volatility. 

This paper aims at studying the link between institutional investments and crypto markets behaviour in a systematic and quantitative fashion, exploiting a novel combination of data sources on a larger sample of cryptocurrencies. 
Our analysis exploits network science tools to study the structure and evolution of the co-investment network, where two cryptocurrencies are linked if they share a common investor. 
The article is organised as follows: in Section \ref{sec:data}, we describe how the data was collected and integrated; in Section \ref{sec:methods}, we present the methodologies and algorithms employed for this study; in Section \ref{sec:network}, we describe the co-investment network and study how the cryptocurrency features (e.g., type of blockchain protocol, use case) are related to the network structure; in Section \ref{sec:correlation} we study the connection between the structure of the co-investment network and market properties of different assets. In Section \ref{sec:discussion} we conclude.

\section{Dataset and methods} % (fold)
\label{sec:data}

\subsection*{Data Description}
In this paper, we use three main data types, namely (i) cryptocurrency price time series data, (ii) cryptocurrency metadata describing projects' technological features and/or their use case and functionalities, and (iii) data capturing information on investment rounds in cryptocurrency projects.

Market data (i) and cryptocurrency metadata (ii) were extracted from the website Coinmarketcap \cite{coinmarket}.
Data covers $1324$ cryptocurrency projects over a period of 8 years, spanning from 2014 to 2022. 
Market data consists of each cryptocurrency's opening price, closing price, and traded volume, sampled weekly.

Coinmarketcap also assigns tags describing the main features of the different cryptocurrencies. 
Metadata can be broadly classified into \emph{technology}-related specifications (e.g., Proof-of-Work vs. Proof-of-staked based blockchain), \emph{ecosystem}-related information (e.g., the cryptocurrency is built on top of a specific blockchain, is or is not defined as a DeFi project), and information on the \emph{use case} (e.g., the cryptocurrency supports a distributed storage project, the cryptocurrency is a fan-token, or it is simply used as a store of value), see Appendix \ref{app:tags} for the list of available tags and their respective frequency. The dataset contains $226$ unique tags. Coinmarketcap also provides cryptocurrencies' webpage URLs, which are used to merge market data with investment data.

Finally, the investments' data (iii) is gathered from Crunchbase \cite{OECD_crunchbase}, a commercial database covering worldwide innovative companies and accessed by 75M users each year. 
The data is sourced through two main channels: an extensive investor network and community contributors. 
Investors commit to keeping their portfolios updated to get free access to the dataset, while 600,000+ executives, entrepreneurs, and investors update over 100,000 company, people, and investor profiles per month. 
Crunchbase processes the data with machine learning algorithms to ensure accuracy and scan for anomalies, ultimately verified by a team of data experts at Crunchbase. 
Due to its broad coverage, the data has been used in thousands of scholarly articles and technical reports \cite{OECD_crunchbase, den2020crunchbase}. 
Information on Crunchbase includes an overview of the company's activities, number of employees, and detailed information on funding rounds, including investors and - more rarely - amounts raised. 
We provide detailed information on the features contained in this dataset in Appendix \ref{app: crunchbase}.

We merged the Crunchbase data on investment rounds with Coinmarketcap data via the companies' webpage URLs. 
After merging, the dataset includes $4395$ investments made in $1458$ rounds by $1767$ investors to $1324$ cryptocurrency projects appearing on Crunchbase. 
The total investments amount to $\$13B$ US dollars in the period considered (2008-2022). When merging with the time series data, we can still track $624$ cryptocurrencies projects.

\subsection*{Methods}
\label{sec:methods}
In this section, we review the methods used for our analyses.
First, we describe how we clustered the network nodes, and then how we analysed the interplay between the network structure and the market dynamics.

\paragraph{Clustering algorithm}
\label{par:clustering_algorithm}
We assign a vector $\mathbf{x}_i$ to each cryptocurrency, where $x_{i,j} = 1$ if the $j$-th tag (see Table \ref{tab:tags_tab} for the list of cryptocurrency tags) is assigned to the $i$-th currency, and $x_{i,j} = 0$ otherwise. 
We used the Ward Aggregative Clustering \cite{Ward1963} algorithm to divide the currencies into different clusters based on the observations $\left(\mathbf{x}_1, \mathbf{x}_2, \ldots, \mathbf{x}_n\right)$. 
The algorithm has a "bottom-up" approach: each observation is initially placed in its own clusters, and clusters are merged sequentially according to some criterion until the desired number of clusters is reached. 
Wards' algorithm specifically prescribes to merge, at each iteration, the pair of clusters $S_i$, $S_j$ that minimizes the distance $\Delta\left(S_i, S_j\right)$, defined as

\begin{equation}
\label{eq:Ward's distance}
    \Delta\left(S_i, S_j\right) = \sum_{l \in S_i \cup S_j } \|\mathbf{x}_l - \boldsymbol{\mu}_{i+j}\|^2 - \sum_{l \in S_i}\|\mathbf{x}_l - \boldsymbol{\mu}_{i}\|^2 - \sum_{l \in S_j}\|\mathbf{x}_l - \boldsymbol{\mu}_{j}\|^2 = \frac{\left|S_i\right|\left|S_j\right|}{\left|S_i\right| + \left|S_j\right|}\|\boldsymbol{\mu}_i - \boldsymbol{\mu}_j\|^2,
\end{equation}

where $\left|S_i\right|$ is the number of observations in cluster $S_i$, $\boldsymbol{\mu}_i$ is the mean of points in $S_i$, $\boldsymbol{\mu}_j$ is the mean of points in $S_j$, and $\boldsymbol{\mu}_{i+j}$ is the mean of points in $S_i \cup S_j$. It should be noted that the number of clusters $k$ is an input of the clustering algorithm. 
Using the elbow method (see Appendix \ref{app:elbow_method}), we set $k=12$.
\paragraph{Clustering evaluation and benchmarks} We investigated whether the clusters obtained with the previous procedure reflect the underlying network structure by studying the in-density and out-density of links according to the partitioning defined by the clusters. 
Given the the adjacency matrix $A$ of our co-investment network and the clustering $S^*=\left\{S_1,\ldots,S_k\right\}$, we define the \textit{in-density} of a cluster $S_i$ as
\begin{equation}
\label{eq:in_cluster}
\rho^i_i = \frac{1}{\left|S_i\right|\left(\left|S_i\right|-1\right)}\sum_{j,k \in S_i, j \neq k} A_{jk},
\end{equation}
and its \textit{out-density} as
\begin{equation}
\label{eq:out_cluster}
\rho^o_i = \frac{1}{\left|S_i\right|\left||S_i\right|}\sum_{j \in S_i, k \notin S_i} A_{jk}.
\end{equation}

We compare the \textit{in-densities} and \textit{out-densities} of the clusters identified by the clustering algorithm with those of random clusters. 
To generate the random clusters, we simply assign each cryptocurrency to one of the twelve possible clusters with equal probability.

\paragraph{Time series processing} The investigation of the co-investment network's impact on the cryptocurrency market involves computing cryptocurrencies' returns correlation. 
The primary objects of this analysis are cryptocurrencies' closing price time series $p_i\left(t\right),\ i=1,\ldots,N$. 
We compute their log returns as
\begin{equation}
\label{eq:log_returns}
r_i\left(t\right) = \log\left(\frac{p_i\left(t+1\right)}{p_i\left(t\right)}\right),
\end{equation}
and use the \textit{leave-one-out} rescaling described in \cite{Bouchaud2003} to define the rescaled returns,
\begin{equation}
\label{eq:rescaled_log_returns}
    \tilde{r}_i(t) = \frac{r_i(t) - \mathbb{E}_{t'}\left[r_i(t')\right]}{\sqrt{\mathbb{V}_{t'\neq t}\left[r_i(t')\right]}}\
\end{equation}
where the average is computed over all times $t'$, but the variance is computed from the time series where the observation corresponding to $t'=t$ has been removed. 
The correlation matrix of the time series $\tilde{r}_i$ is defined as
\begin{equation}
\label{eq:correlation_matrix}
 C_{ij} = \mathbb{E}_t\left[r_i\left(t\right)r_j\left(t\right)\right].
\end{equation}
Cryptocurrencies' prices usually move coherently, increasing or decreasing simultaneously \cite{katsiampa2019high,koutmos2018return,stosic2018collective}. 
This collective behavior of the market makes returns strongly correlated and hides the more subtle relationships we want to highlight. 
We adopt the following strategy to remove the so-called {\em market component} from the correlation matrix \cite{laloux1999noise}. 
We first compute the set of eigenvalues $\lambda_1, \ldots, \lambda_N$ of the correlation matrix, the corresponding eigenvectors $\mathbf{v}_1, \ldots, \mathbf{v}_N$, and the modes $m_i\left(t\right)$, defined as
\begin{equation}
\label{eq:collective_modes}
m_i\left(t\right) = \sum_{j=1}^N v_{ij}\tilde{r}_j\left(t\right).
\end{equation}
We call \textit{market mode} the mode $m_1\left(t\right)$ associated with the largest eigenvalue $\lambda_1$. The time series $\tilde{r}_i\left(t\right)$ can now be written as linear combinations of the modes $m_i\left(t\right)$,
\begin{equation}
\label{eq:time_series_as_combinations_of_collective_modes}
\tilde{r}_i\left(t\right) = \sum_{j=1}^N c_{ij}m_{j}\left(t\right).
\end{equation}
We can now define the {\em adjusted} time series $r'_i\left(t\right)$,
\begin{equation}
\label{eq:clean_time_series}
r'\left(t\right) = \sum_{j>1}^N c_{ij}m_{j}\left(t\right),
\end{equation}
and the {\em adjusted} correlation matrix $C'$,
\begin{equation}
\label{eq:clean_correlation_matrix}
C'_{ij} = \mathbb{E}_t\left[r'_i\left(t\right)r'_j\left(t\right)\right].
\end{equation}

\paragraph{Network correlation and random benchmarks} We compute the average value of the matrix $C$ and $C'$ across the pairs of cryptocurrencies $\left(i, j\right)$ that share a link in the co-investment network. Given any (binary) adjacency matrix $\mathbf{M}$ we define
\begin{equation}
    C_{\mathbf{M}} = \mathbb{E}_{ij} \left[C_{ij}M_{ij}\vert M_{ij}>0\right],
    \label{eq:corr1}
\end{equation}
and
\begin{equation}
    C'_{\mathbf{M}} = \mathbb{E}_{ij} \left[C'_{ij}M_{ij}\vert M_{ij}>0\right],
    \label{eq:corr2}
\end{equation}
where the average runs over all pairs $\left(i,j\right)$ of connected nodes. 
We compute $C_{\mathbf{A}}$ and $C'_{\mathbf{A}}$ over the adjacency matrix $A$ of the real co-investment network, and compare them with the values obtained on three random network models: the \textit{Erd\H{o}s-R\'enyi} model\cite{ErdosRenyi}, the \textit{Stochastic Block Model}\cite{KarrerNewman2011}, and the \textit{Configuration Model}\cite{Newman2003}. 
For every model, we sample $n=1000$ network instances $R_1, \ldots, R_n$ at random, and compute the mean and standard deviation of the sets $\left\{C_{\mathbf{R}_1}, \ldots, C_{\mathbf{R}_n}\right\}$ and $\left\{C'_{\mathbf{R}_1}, \ldots, C'_{\mathbf{R}_n}\right\}$. 
All models are parametrized to match the empirical properties of the co-investment network. The probability of a link $p$ in the Erd\H{o}s-R\'enyi model is set to match the co-investment network's empirical density, 
$$
p = \frac{1}{N\left(N-1\right)}\sum_{i=1}^N\sum_{j>i}^N A_{ij}.
$$
Blocks in the Stochastic block model match the clusters found with the clustering algorithm and the densities within- and across- clusters are equal to the empirical values, respectively $\rho^i_1, \ldots, \rho^i_k$, and $\rho^o_1, \ldots, \rho^o_k$. Finally, the degree sequence in the configuration model matches the empirical degree sequence.
\begin{figure}[h!]
    \centering
    \includegraphics[width=.6\textwidth]{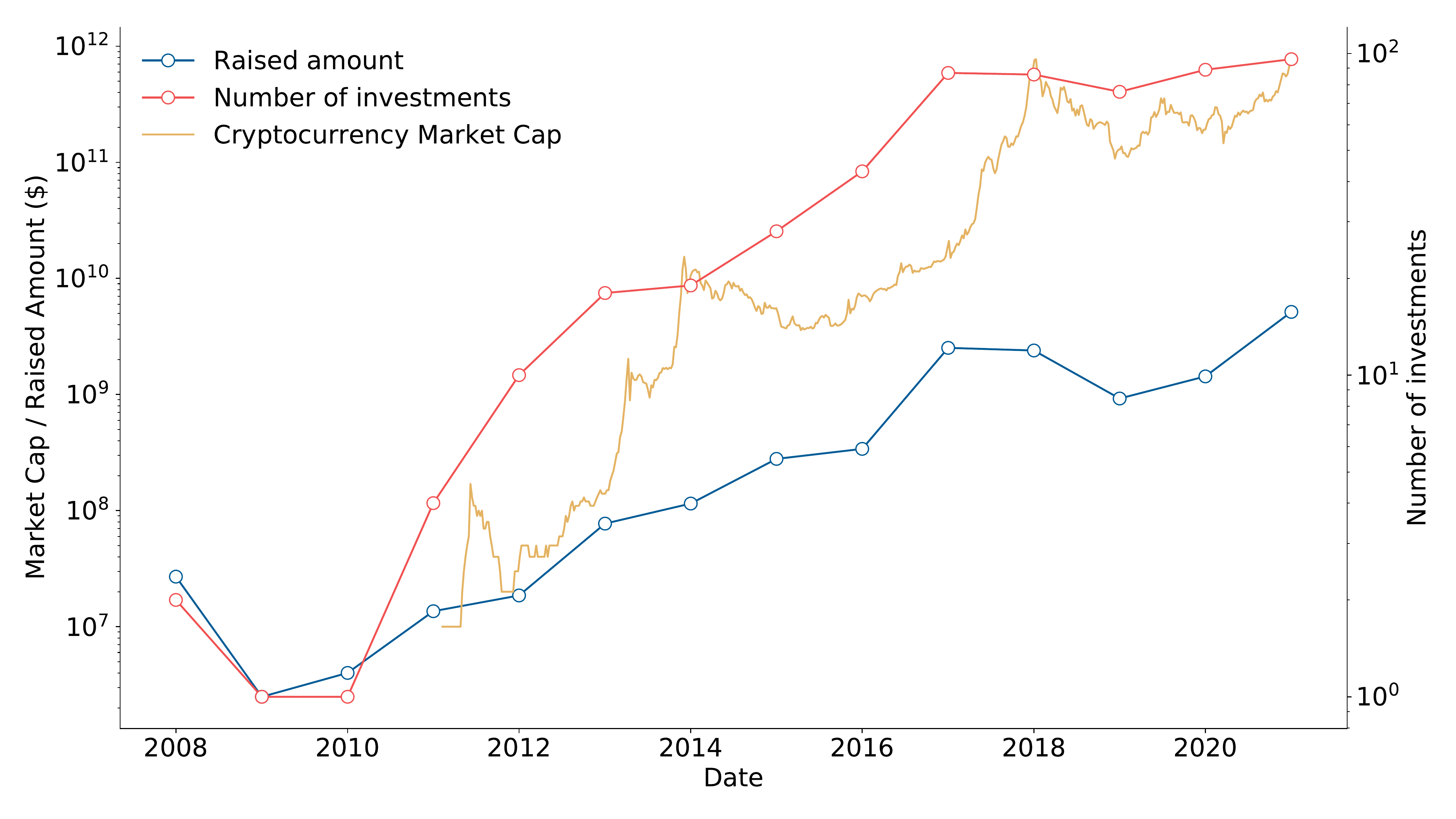}
    \caption{\textbf{Temporal evolution of institutional investments in cryptocurrency projects.} Yearly total amount raised in USD (blue line) and the number of investments (red line) in cryptocurrency projects retrieved from the Crunchbase dataset for the period 2009-2012. The total capitalization of the cryptocurrency market in USD is shown in yellow.}
    \label{fig:investments_time}
\end{figure}

\section{Results}
\label{sec:results}

\subsection{Structure of the cryptocurrency co-investment network}
\label{sec:network}

In this section, we analyze the relationship between institutional investments and cryptocurrency market response.
We start by quantifying the joint evolution of the number and volume of investments and the overall growth of the cryptocurrency market. In Fig. \ref{fig:investments_time}, we show for the evolution in time of the total raised amount, number of investments and market capitalisation of the cryptocurrency ecosystem. In Sec. \ref{sec:data} we discuss the details on how the investment rounds and amounts were retrieved from Crunchbase.
Overall, we find that the number of investments, as well as the amount raised, has been steadily growing since 2012. Moreover, we found a positive correlation between the cryptocurrency market cap (MC) and both the total volume of investments (VI) and the number of investments (NI). The Spearman correlation amounts respectively to $\rho_{MC-VI}=0.79$  and $\rho_{MC-NI}=0.81$.

\begin{figure}[ht!]
	\framebox{\includegraphics[width = \textwidth]{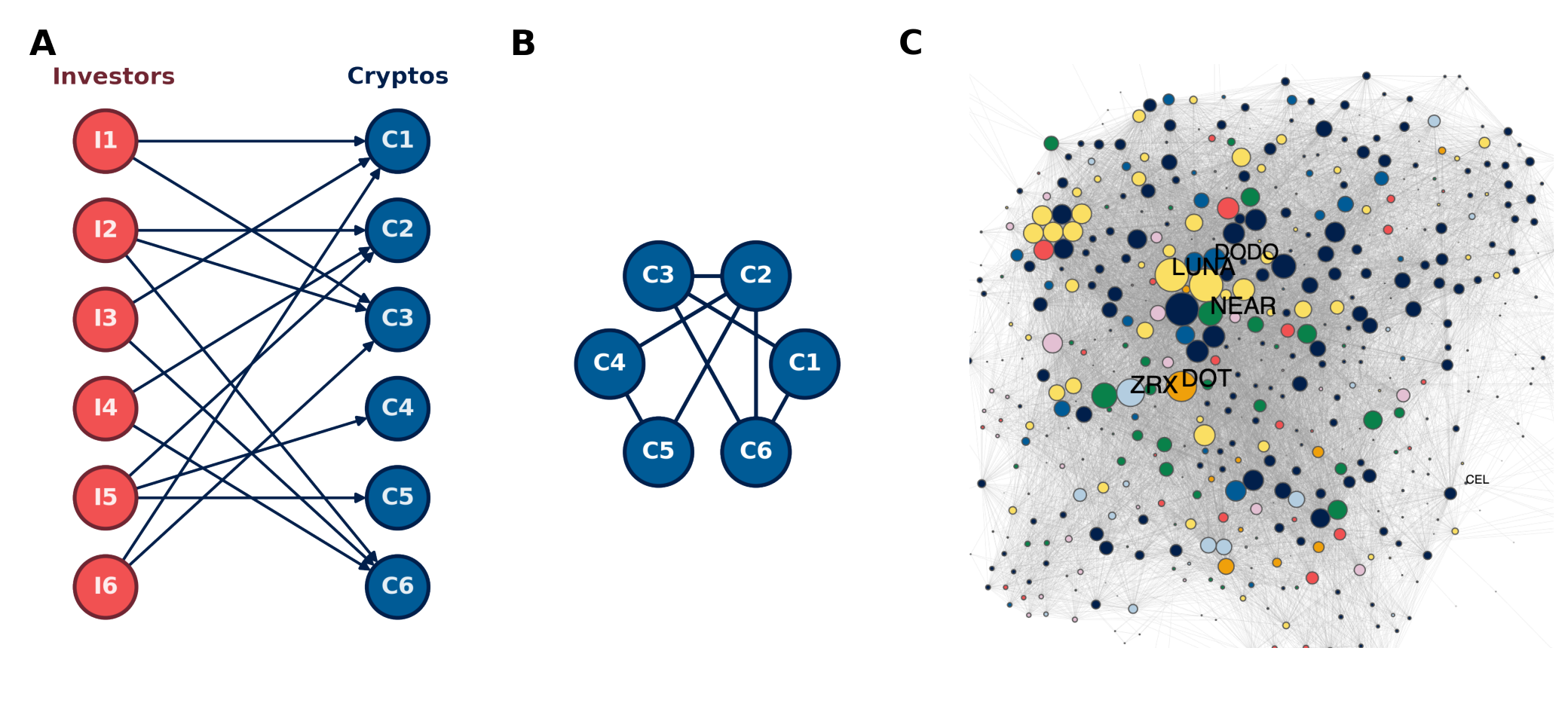}}
	\caption{\textbf{Cryptocurrencies co-investment network.} (A) The Crunchbase dataset can be mapped into a bipartite network where investors are connected to cryptocurrency projects they have invested in at least once. We use an approach similar to Lucchini et al., 2020 \ref{lucchini_et_al_2020} (B) Projection of the bipartite investors-cryptocurrencies network, where two cryptocurrencies are linked if they have at least a common investor. (C) Real co-investment network of $624$ cryptocurrency projects with at least one connection. Node size is proportional to the number of connections, and link width is proportional to the number of common investors between two cryptocurrencies. Colors represent different groups of cryptocurrencies clustered according to their tags' similarity on Coinmarket cap. We report the name of the top node by degree in five representative clusters (DODO, LUNA, NEAR, ZRX, DOT).}
	\label{fig:c2cnetwork}
\end{figure}

To better characterize investment patterns in the cryptocurrency ecosystem, we analyse the cryptocurrency co-investment network, where two projects are connected if they share an investor. 
In Fig. \ref{fig:c2cnetwork}, we schematically show how the co-investment network is constructed as a monopartite projection of the bipartite network where investors are connected to cryptocurrency projects they have funded at least once (panels A-B). The methods is similar to the one developed by Lucchini et. al, 2020 \ref{lucchini_et_al_2020}. In Fig. \ref{fig:c2cnetwork}C, we show the real co-investment network composed of $624$ cryptocurrency projects. The node sizes are proportional to their degree, and the link widths are proportional to the number of common investors between two cryptocurrencies.

To capture the relationship between the network structure and the characteristics of cryptocurrencies, we assign cryptocurrencies to clusters based on their properties, e.g., technological features and use case. In particular, we use an agglomerative clustering algorithm using as features the Coinmarketcap tags (described in Sec. \ref{sec:data} and Tab. \ref{tab:tags_tab}): two projects are more likely to be assigned to the same cluster the more similar their tags are. Clusters are depicted in different colors in panel C in Fig. \ref{fig:c2cnetwork}.

\begin{figure} 
    \centering
    \framebox{\includegraphics[width=.9\textwidth]{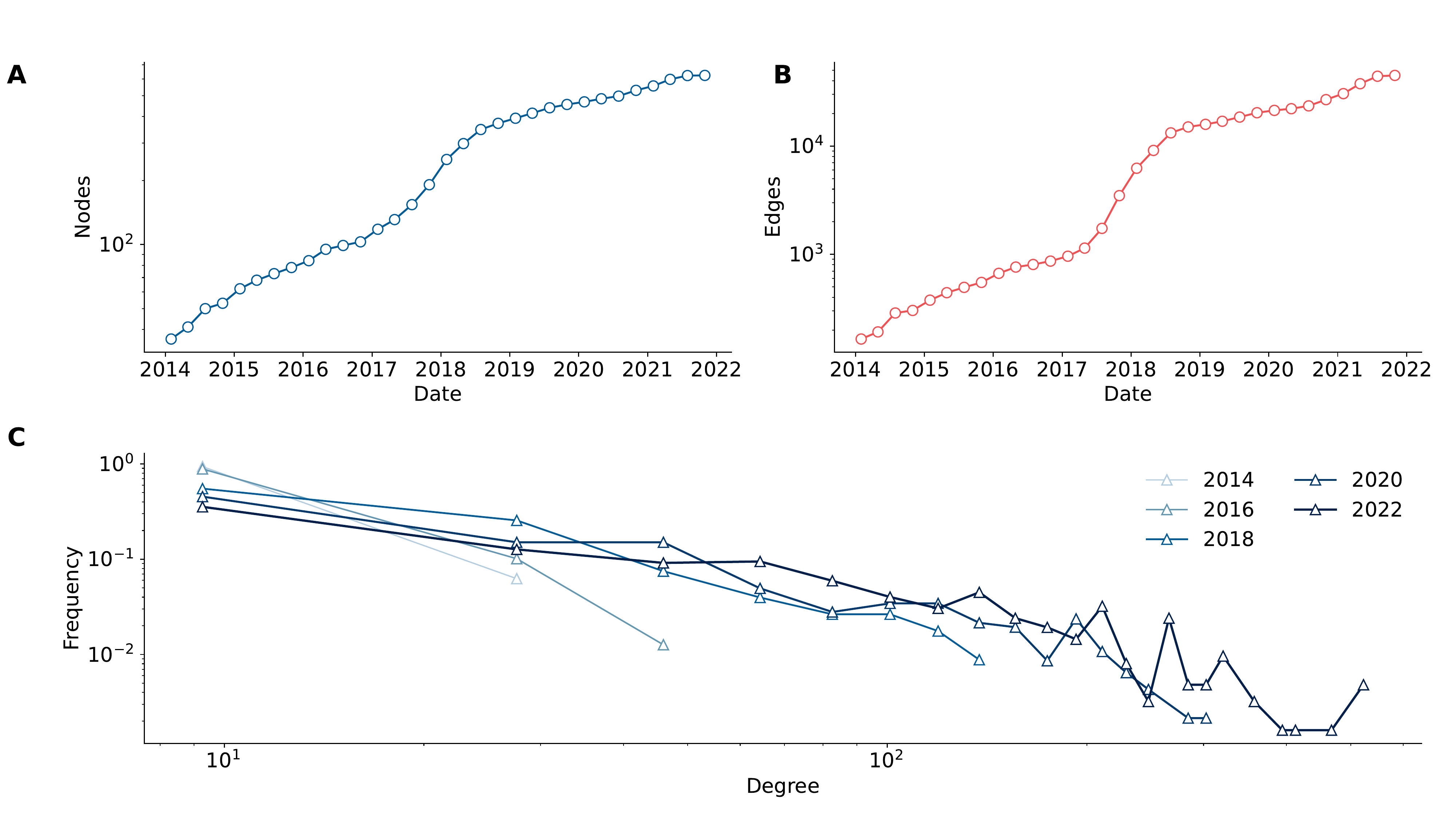}}
    \caption{{\bf Time evolution of network metrics.} In Panel A we report the cumulative number of nodes in the co-investment network. Panel B represents the cumulative number of edges, i.e. new investors supporting cryptocurrency projects. In Panel C we plot the degree distribution for five representative years.}
    \label{fig:metrics_evolution}
\end{figure}

Figure \ref{fig:metrics_evolution} shows the evolution of the co-investment network in time.  

We found that, since 2014, the network has grown steadily in terms of the cumulative number of nodes (panel A), i.e., cryptocurrency projects funded by institutional investors, and cumulative number of edges (panel B), i.e., common investors between cryptocurrencies. %This is consistent with our finding that the number of investment rounds has been increasing since 2012 (Fig. \ref{fig:c2cnetwork}). 
Interestingly, the growth displays a steeper increase around 2017-2019, consistently with the rapid increase in demand for cryptocurrencies and Bitcoin's valuation over those years \cite{coinreport2018}.

We also find that the degree distribution characterizing the number of connections per node in the co-investment network is heavy-tailed, with most nodes having a single connection and only a few having hundreds of neighbors (see Fig. \ref{fig:c2cnetwork}C). The shape of the distribution has been relatively stable over time,  as shown by plotting the distribution for five representative years in the chosen time frame. This observation is also compatible with the findings discussed in \cite{liu2021}, where the co-investment network for ERC-20 tokens only exhibits similar structural properties.

\begin{figure}[htb!]
    \centering
    \includegraphics[width=.8\textwidth]{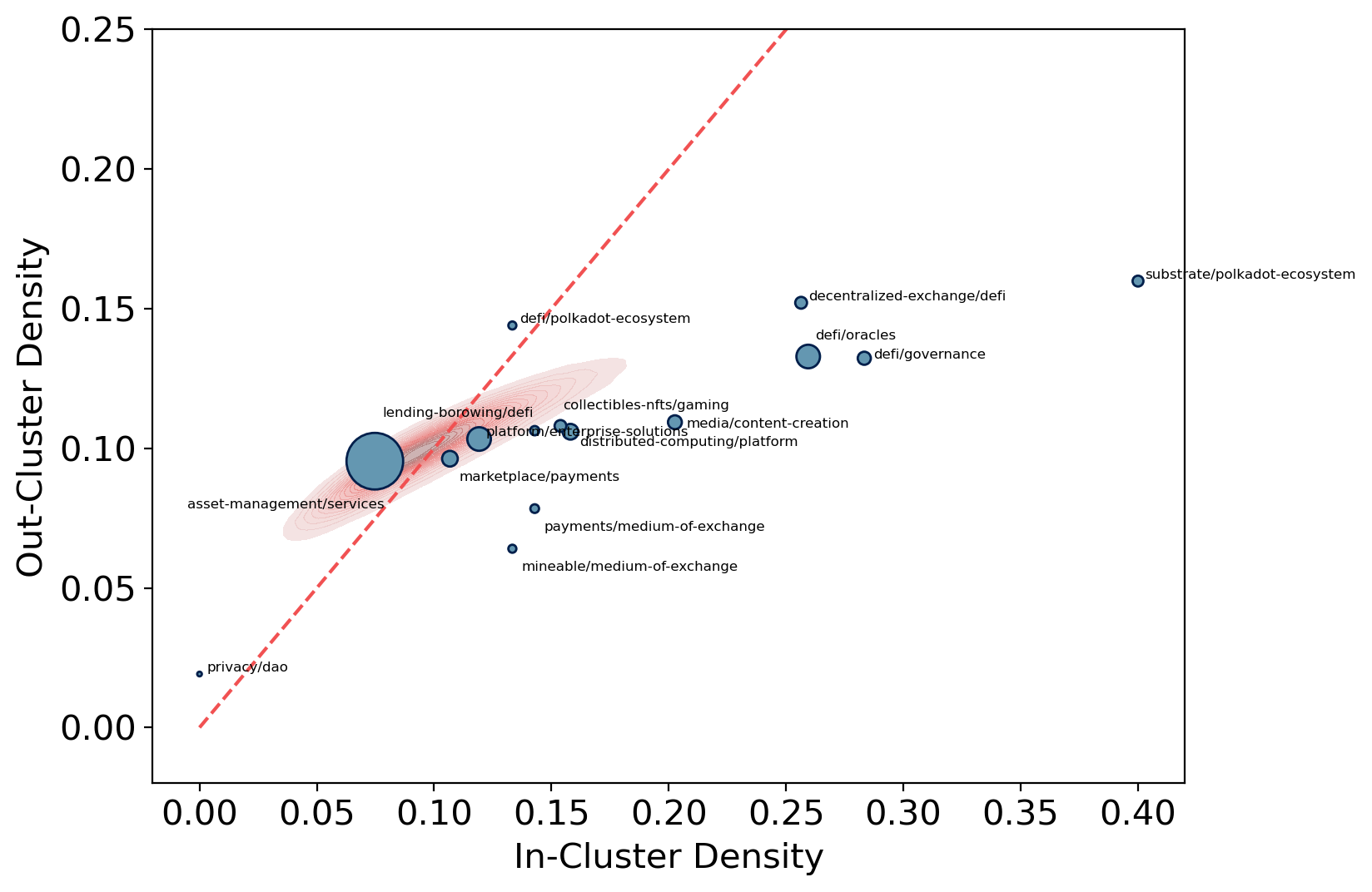}
    \caption{{\bf Comparison of in- and out- cluster densities.} In- and out-densities measured on 15 clusters generated by running the clustering algorithm on the cryptocurrencies' tags. Blue circles represents the different clusters (the size of the circle is related to the cluster's size) and the text indicates the most relevant tags per cluster. The dashed red line is the diagonal, the red-shaded area represents the in- and out-cluster density distribution for the randomised network.}
    \label{fig:in_and_out_density}
\end{figure}

We then study the relationship between the underlying co-investment network's structure and the cryptocurrency features. We first calculate in- and out- cluster densities as defined in Eq.\eqref{eq:in_cluster} and Eq. \eqref{eq:out_cluster} respectively on the real clusters and on randomised ones (as explained in Sec. \ref{sec:data}).
Figure \ref{fig:in_and_out_density} shows that the densities inside clusters of similar cryptocurrencies are usually larger than those across clusters. Comparing our results with those obtained for random clusters (red shaded area in Fig. \ref{fig:in_and_out_density}), we find significant differences. This means that the more cyrptocurrency projects are similar, the larger number of investors they tend to have in common.
This result suggests a connection between the topology of the network and the intrinsic features of cryptocurrency projects and hints at the presence of specialised investors that do not simply invest in the whole cryptocurrency ecosystem at random but focus on specific technologies and/or use cases.

\subsection{Interplay between the co-investment network structure and returns correlations}
\label{sec:correlation}

In this section, we investigate the interplay between the structure of the co-investment network and the cryptocurrency market properties. 
More specifically, we examine whether the presence of a shared investor tends to anchor cryptocurrencies' prices to one another.

We, therefore, compute average returns correlation $C_{\textbf{A}}$ defined in Eq.\eqref{eq:corr1} across pairs of cryptocurrencies sharing a link in the real co-investment network (described by its adjacency matrix $A$). We also compute average returns correlation $C'_{\bf{\tilde{A}}}$ of cryptocurrency pairs sharing a link on random network benchmarks including (i) an Erdos-Renyi network, (ii) a configuration model and (iii) a stochastic block model parametrized to reproduce some of the features of the real network (e.g., number of nodes, number of clusters, degree distribution - as detailed in Sec. \ref{sec:data}). Fig. \ref{fig:correlation_panel} compares the values of $C_{\mathbf{A}}$ and $C'_{\mathbf{\tilde{A}}}$ for the real co-investment network and the benchmarks respectively. 
We, first, monitor the average returns correlation between cryptocurrency pairs conditioned on their distance in the co-investment network and the benchmark networks. As shown in Fig. \ref{fig:correlation_panel}, Panel A the returns correlation between the closest cryptocurrency pairs (at distance $d=1,2$) in the real co-investment network is higher compared to all random benchmarks. In this panel, all the correlations were rescaled by the correlation of the real network. This shows that cryptocurrency pairs that are more closely related in the co-investment network, have also more connected market dynamics.

Fig. \ref{fig:correlation_panel}, Panel B shows the average returns correlation for the real network (dark blue) and all benchmarks (rescaled by the correlation of the real network). The average correlation on the real network is significantly larger than all the benchmarks tested, suggesting that the network's structure may directly impacts the cryptocurrencies' market behaviour. 

This is even more evident when we remove the market component from the returns correlation (light blue, green, red and orange bars in Fig. \ref{fig:correlation_panel} Panel B representing respectively the rescaled correlation of the real network and the three random benchmarks, namely the configuration model, stochastic block model and Erd\H{o}s-R\'eny) as discussed in Sec. \ref{sec:methods}. In particular, the returns correlation in the real network is not affected when considering $C'_{A}$ (light blue bar).

\begin{figure}[htb!]
    \centering
   	\includegraphics[width = .70\textwidth]{figures/correlations_panel.pdf}
    \caption{{\bf Returns correlation of connected cryptocurrency pairs.}
    Panel A: Average correlation of returns time series on the real network and random benchmarks as a function of the network distance. The $y$-axis is rescaled by the correlation value of the co-investment network. Panel B: Average correlation for cryptocurrencies connected in the co-investment network $C_{\mathbf{A}}$ (blue bars) and in random benchmarks (red - configuration model, green - stochastic block model,  orange - Erd\H{o}s-R\'enyi). For each network, the bottom bar shows the correlation obtained after removing the market component ($C'_{\mathbf{A}}$). Correlation values have been rescaled between $[0,1]$ for visual clarity.}
    \label{fig:correlation_panel}
\end{figure}

\section{Discussion}
\label{sec:discussion}
In this paper, we have analyzed an ecosystem of $1324$ cryptocurrency projects that received $4395$ investments from $1767$ investors for a total amount of \$13B appearing on Crunchbase. We have built and analysed the co-investment network, where two cryptocurrencies are linked if they share an investor. We have also clustered cryptocurrency projects based on metadata and tags from the Coinmarketcap website and studied the community structure.

As hinted by previous research and surveys concerning institutional and individual crypto investors preferences  \cite{fidelityreport, OECD, ciaian2016economics, liu2021}, our results show that investors tend to specialise and focus on particular technologies, use cases and features of the cryptocurrency projects they decide to include in their portfolio.

We have also analyzed the relation between the co-investment network and the cryptocurrencies' market properties. We showed that the returns of cryptocurrencies sharing a link in the co-investment network tend to be correlated. The marginal increase in correlation of cryptocurrency returns decreases as the distance between the considered pairs of cryptoccurrencies in the co-investment network increases. 

Our work has limitations that, hopefully, can be turned into future avenues of research. Firstly, our data collection process stopped over the summer of 2021, before the second major cryptocurrencies crash and the default of established players such as Terra, Celsius, and FTX. It is legit to wonder to what extent our results would hold in the new regime, where the general sentiment towards cryptocurrencies has pivoted. 

Secondly, some prominent players in the cryptocurrencies ecosystem are not associated with a company, but rather with different types of organizations including \textit{Decentralized Autonomous Organizations} (DAOs), foundations, or even no legal entity at all. The nature of the investment may also vary substantially. For instance, instead of buying a share of the company, investors may, e.g., lend money to DeFi protocols in exchange for tokens as rewards (a practice known as \textit{liquidity mining}\cite{fan2022towards}). These new organization types and forms of investment are scarcely represented in our dataset, therefore we can only offer a partial view of the cryptocurrencies investment ecosystem.  Finally, most of our analysis was performed on a static network. However, how the network grows, what the different investment strategies adopted by an investor are, and how they depend on the market are also clearly worth analyzing.

In light of the recent crypto marker crash events - from the stablecoin pair Terra -- Luna to large exchanges \cite{briola2022anatomy,hermans2022decrypting,cryptocontagion} -- undestanding the crypto market connectedness at the investors level help shed light on possible contagion channels posing threat to the ecosystem overall stability.

\bibliographystyle{unsrt}
\bibliography{biblio.bib}

\newpage 

\begin{center}
\Huge{\textbf{Appendix }}
\end{center}

\appendix

% Appendix starts here

\section{Coinmarketcap cryptocurrency tags}
\label{app:tags}
We include below a table containing all the tags together with their respective frequency gathered from Coinmarketcap for all the cryptocurrency projects analysed in this paper. 

\begin{longtable}[!h]{llll}
    \\0 & mineable: 465 & defi: 333 & platform: 188 \\1 & collectibles-nfts: 139 & yield-farming: 129 & payments: 127 \\2 & pow: 98 & marketplace: 97 & binance-smart-chain: 86 \\3 & masternodes: 84 & decentralized-exchange: 83 & smart-contracts: 82 \\4 & exnetwork-capital-portfolio: 72 & hybrid-pow-pos: 72 & medium-of-exchange: 65 \\5 & polkadot-ecosystem: 53 & governance: 53 & scrypt: 53 \\6 & dao: 49 & enterprise-solutions: 47 & ethereum: 47 \\7 & privacy: 42 & gaming: 41 & media: 40 \\8 & pos: 38 & asset-management: 37 & kinetic-capital: 36 \\9 & stablecoin: 32 & centralized-exchange: 32 & distributed-computing: 31 \\10 & services: 28 & ai-big-data: 28 & content-creation: 27 \\11 & cosmos-ecosystem: 26 & staking: 26 & iot: 26 \\12 & pantera-capital-portfolio: 23 & alameda-research-portfolio: 23 & filesharing: 23 \\13 & tokenized-stock: 22 & sha-256: 22 & substrate: 22 \\14 & polkastarter: 20 & amm: 20 & memes: 19 \\15 & sports: 18 & gambling: 18 & derivatives: 18 \\16 & storage: 17 & x11: 16 & oracles: 16 \\17 & rebase: 16 & solana-ecosystem: 16 & stablecoin-asset-backed: 16 \\18 & entertainment: 15 & store-of-value: 14 & polkadot: 14 \\19 & yield-aggregator: 14 & wallet: 14 & dao-maker: 14 \\20 & coinbase-ventures-portfolio: 13 & duckstarter: 13 & binance-launchpad: 13 \\21 & wrapped-tokens: 12 & seigniorage: 12 & interoperability: 12 \\22 & lending-borowing: 10 & binance-chain: 10 & cms-holdings-portfolio: 10 \\23 & dapp: 10 & insurance: 10 & dcg-portfolio: 9 \\24 & multicoin-capital-portfolio: 9 & launchpad: 9 & polychain-capital-portfolio: 9 \\25 & hashkey-capital-portfolio: 9 & fan-token: 9 & synthetics: 8 \\26 & poolz-finance: 8 & binance-labs-portfolio: 8 & three-arrows-capital-portfolio: 8 \\27 & placeholder-ventures-portfolio: 7 & blockchain-capital-portfolio: 6 & scaling: 6 \\28 & social-money: 6 & fabric-ventures-portfolio: 6 & crowdfunding: 6 \\29 & dpos: 5 & boostvc-portfolio: 5 & arrington-xrp-capital: 5 \\30 & framework-ventures: 4 & defi-index: 4 & trustswap-launchpad: 4 \\31 & discount-token: 4 & state-channels: 3 & coinfund-portfolio: 3 \\32 & logistics: 3 & dex: 3 & a16z-portfolio: 3 \\33 & marketing: 3 & e-commerce: 3 & tourism: 3 \\34 & health: 2 & research: 2 & loyalty: 2 \\35 & dragonfly-capital-portfolio: 2 & identity: 2 & energy: 2 \\36 & parafi-capital: 1 & huobi-capital: 1 & metaverse: 1 \\37 & yearn-partnerships: 1 & defiance-capital: 1 & ledgerprime-portfolio: 1 \\38 & data-provenance: 1 & sharing-economy: 1 & zero-knowledge-proofs: 1 \\39 & paradigm-xzy-screener: 1 & electric-capital-portfolio: 1 & 1confirmation-portfolio: 1 \\40 & binance-launchpool: 1 & video: 1 & analytics: 1 \\41 & music: 1 & cybersecurity: 1 & prediction-markets: 1 \\42 & fenbushi-capital-portfolio: 1 & options: 1 & education: 1 \\43 & real-estate: 1 & x13: 1 & aave-tokens: 1 \\44 & avalanche-ecosystem: 1 & mobile: 1 & galaxy-digital-portfolio: 1 \\45 & crowdsourcing: 1 & hardware: 0 & reputation: 0 \\46 & usv-portfolio: 0 & jobs: 0 & stablecoin-algorithmically-stabilized: 0 \\47 & quark: 0 & multiple-algorithms: 0 & equihash: 0 \\48 & events: 0 & winklevoss-capital: 0 & art: 0 \\49 & atomic-swaps: 0 & cryptonight: 0 & communications-social-media: 0 \\50 & neoscrypt: 0 & social-token: 0 & dag: 0 \\51 & heco: 0 & retail: 0 & eth-2-0-staking: 0 \\52 & philanthropy: 0 & commodities: 0 & ringct: 0 \\53 & transport: 0 & sharding: 0 & quantum-resistant: 0 \\54 & ethash: 0 & vr-ar: 0 & hospitality: 0 \\55 & asset-backed-coin: 0 & layer-2: 0 & blake2b: 0 \\56 & hybrid-dpow-pow: 0 & hacken-foundation: 0 & adult: 0 \\57 & manufacturing: 0 & sha-256d: 0 & search-engine: 0 \\58 & ontology: 0 & dagger-hashimoto: 0 & poc: 0 \\59 & pos-30: 0 & blake256: 0 & blake: 0 \\60 & hybrid-pos-lpos: 0 & geospatial-services: 0 & m7-pow: 0 \\61 & fashion: 0 & cryptonight-lite: 0 & tron: 0 \\62 & mimble-wimble: 0 & lp-tokens: 0 & poi: 0 \\63 & lyra2rev2: 0 & agriculture: 0 & posign: 0 \\64 & timestamping: 0 & pop: 0 & lpos: 0 \\65 & sidechain: 0 & platform-token: 0 & eos: 0 \\66 & hybrid-pow-npos: 0 & lelantusmw: 0 & groestl: 0 \\67 & cosmos: 0 & x11gost: 0 & scrypt-n: 0 \\68 & food-beverage: 0 & tpos: 0 & qubit: 0 \\69 & x15: 0 & sha-512: 0 & data-availability-proof: 0 \\70 & cuckoo-cycle: 0 & escrow: 0 & rollups: 0 \\71 & hybrid-pos-pop: 0 & yescript: 0 & rpos: 0 \\72 & x14: 0 & post: 0 & blake2s: 0 \\73 & nist5: 0 & bulletproofs: 0 & sigma: 0 \\74 & argon2: 0 & lyra2re: 0 & xevan: 0 \\75 & waves: 0 & &  \\
    \caption{Coinmarketcap cryptocurrencies tags and their frequency among the cryptocurrencies in the coinvestment network.}
    \label{tab:tags_tab}
\end{longtable}

\section{Crunchbase dataset}
\label{app: crunchbase}
Crunchbase provides information on worldwide innovative companies. The dataset covers several aspects of the companies, spanning from a basic description of the business description to their financial status, board composition, and even media exposition. The dataset is organized in different bundles that reflect this different information. The bundles are: 
\begin{itemize}
    \item \textbf{Company-related:} \textit{organizations} (including information on parent companies, organization descriptions, and their division in categories) and \textit{investment funds},
    \item \textbf{Investment-related}: \textit{funding rounds} (group of investments in a single company), \textit{investments} (specific investor-to-company transaction), \textit{investors}, \textit{acquisitions}, \textit{ipos},
    \item \textbf{People-related}: \textit{people} covered in the dataset, the \textit{jobs} they have, and the \textit{degrees} they hold, with a focus on \textit{investment partners},
    \item \textbf{Event-related}: \textit{events} description and \textit{event appearances} of specific companies.
\end{itemize}

For the sake of this paper, the relevant bundles concern organization, funding rounds, and investments. We detail their content in Tables \ref{tab:data_description1}, \ref{tab:data_description2}, \ref{tab:data_description3}.

\begin{table}[h]
    \centering
    \begin{tabular}{lrl}
    \toprule
    Bundle   &  Columns Name          & Description \\
    \midrule
    Organization & uuid &  Organization unique Identifier \\
                & name & Company's name\\
                & permalink &             \\
                & cb\_url & Company url on Crunchbase \\
                & rank &   Crunchbase rank\\
                & created\_at &  Record creation date \\
                & updated\_at &  Last record update\\
                & legal\_name &  Company legal name\\
                & roles &   Company, Investor, or both\\
                & domain & Company's website domain\\
                & homepage\_url &  Company's hompage URL\\
                & country\_code &             \\
                & state\_code &             \\
                & region &             \\
                & city &             \\
                & address &             \\
                & postal\_code &             \\
                & status &             \\
                & short\_description &             \\
                & category\_list & \small{Company classification (e.g., Enterprise Software, Financial Services, Social Media)}\\
                & category\_groups\_list & \small{Company classification (e.g., Content and Publishing, Internet Services)}\\
                & num\_funding\_rounds & Number of funding rounds\\
                & total\_funding\_usd & Total Funding raised in USD \\
                & total\_funding &   Total funding raised\\
                & total\_funding\_currency\_code &  Funding currency \\
                & founded\_on &             \\
                & last\_funding\_on &             \\
                & closed\_on &             \\
                & employee\_count &             \\
                & email &             \\
                & phone &             \\
                & facebook\_url &             \\
                & linkedin\_url &             \\
                & twitter\_url &             \\
                & logo\_url &             \\
                & alias1 & Other company's names\\
                & alias2 &             \\
                & alias3 &             \\
                & primary\_role & Either "company" or "investor"\\
                & num\_exits &             \\
    \bottomrule
    \end{tabular}
    \caption{Data entries in the organization Crunchbase bundle.}
    \label{tab:data_description1}
\end{table}

\begin{table}[h]
    \centering
    \begin{tabular}{lll}
    \toprule
    Bundle   &  Columns Name          & Description \\
    \midrule
    Funding Rounds & uuid & Funding round unique identifier\\
                & name & Funding round name (e.g., Angel Round - Facebook)\\
                & permalink &             \\
                & cb\_url & Crunchbase url\\
                & rank & Crunchbase company rank\\
                & created\_at & Record creation date\\
                & updated\_at & Record last update\\
                & country\_code &             \\
                & state\_code &             \\
                & region &             \\
                & city &             \\
                & investment\_type & Investment type (e.g., angel, seed, series a)\\
                & announced\_on &             \\
                & raised\_amount\_usd &             \\
                & raised\_amount &             \\
                & raised\_amount\_currency\_code &             \\
                & post\_money\_valuation\_usd &             \\
                & post\_money\_valuation &             \\
                & post\_money\_valuation\_currency\_code &             \\
                & investor\_count & Number of investors\\
                & org\_uuid & Investee unique identifier\\
                & org\_name & Investee name\\
                & lead\_investor\_uuids & Lead investor's unique identifier.\\
    \bottomrule
    \end{tabular}
    \caption{Data entries in the Crunchbase funding rounds bundle.}
    \label{tab:data_description2}
\end{table}

\begin{table}[h]
    \centering
    \begin{tabular}{lll}
    \toprule
    Bundle   &  Columns Name          & Description \\
    \midrule
    Investments & uuid & Investment unique identifier\\
                & name & Investment's name (e.g., Accel investment in Series A - Facebook)\\
                & permalink &             \\
                & cb\_url & Crunchbase's investment url\\
                & created\_at & Record creation date\\
                & updated\_at & Record last update\\
                & funding\_round\_uuid &             \\
                & funding\_round\_name &             \\
                & investor\_uuid &             \\
                & investor\_name &             \\
                & investor\_type & Either "organization" or "person"\\
                & is\_lead\_investor &             \\
    \bottomrule
    \end{tabular}
    \caption{Data entries in the Crunchbase investment bundle.}
    \label{tab:data_description3}
\end{table}

\section{Methods}
\label{app:methods}
\subsection{Elbow method}
\label{app:elbow_method}
We used the elbow method to choose the total number of cryptocurrencies clusters. For each possible partition $\mathbf{S} = {S_1, \ldots, S_k}$ of the dataset, we first defined a loss function $L\left(S\right)$ as
\begin{equation}
\label{eq:inertia}
    L\left(\mathbf{S}\right) = \sum_{i = 1}^k\sum_{j \in S_{i}} \|\mathbf{x}_j - \boldsymbol{\mu}_i\|^2.
\end{equation}
We ran the clustering algorithm for several different values of $k$, and computed the value of the loss function for the set of optimal partitions $\left\{S^*_{k=1}, S^*_{k=2}, \ldots, S^*_{k=N}\right\}$, where $N$ is the total number of cryptocurrencies in our study.
\begin{figure}[h]
\centering
\includegraphics[width=.8\textwidth]{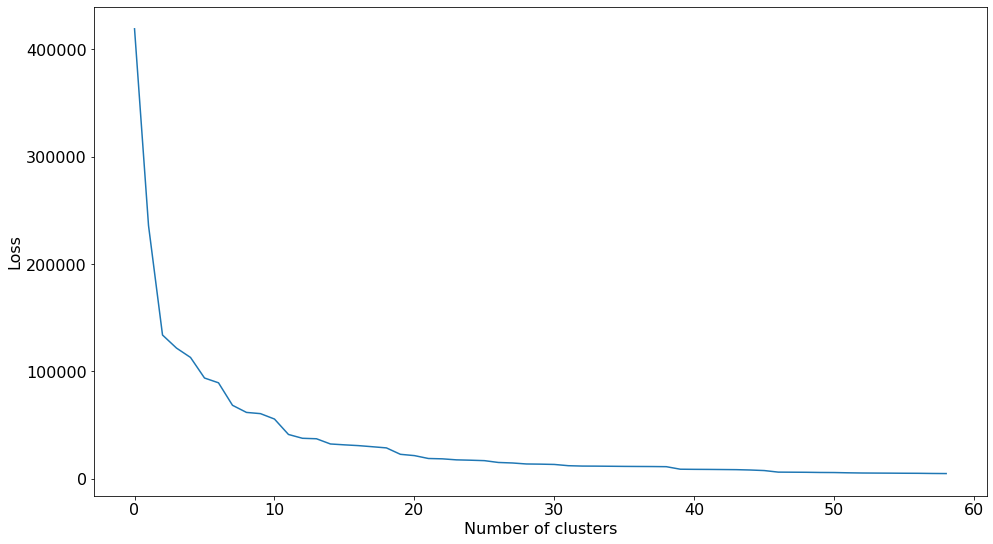}
\caption{Values of the loss function for the different number of clusters. The curve becomes flat when the number of clusters is around $k=12$}
\label{fig:elbow}
\end{figure}
The elbow method prescribes to choose the maximum number of clusters before the curve becomes flat. Intuitively, the method recommends to pick a point where the marginal decrease in the loss function is not worth the additional cost of creating another cluster. Figure \ref{fig:elbow} shows that a value around $k=12$ is compatible with the elbow method in our case.

% Appendix ends here

\end{document}